\documentclass[twocolumn,prb,amsmath,amssymb,superscriptaddress,noeprint]{revtex4-1}
\usepackage[utf8]{inputenc}
\usepackage{hyperref}
\usepackage{amsfonts,amsmath,amssymb}
\usepackage{graphicx}
\usepackage{bm} 
\usepackage{mathrsfs}
\usepackage{subfigure}
\usepackage{textcomp}
\usepackage{hyperref}
\usepackage[flushleft]{threeparttable}
\usepackage[per-mode=symbol]{siunitx}
\usepackage[version=3]{mhchem}

\usepackage{amsmath,amssymb,latexsym}
\usepackage{graphicx}
\usepackage{soul}
\usepackage{bm}
\usepackage{mathrsfs}
\usepackage{physics}
\usepackage{siunitx}
\usepackage{mhchem}
\usepackage[normalem]{ulem}
\usepackage{color}
\usepackage{nicefrac}
\usepackage{todonotes}
\usepackage{multirow}
\usepackage{hyperref}
\usepackage[normalem]{ulem}

\newcommand{\beq}{\begin{equation}}
\newcommand{\eeq}{\end{equation}}
\newcommand{\beqa}{\begin{eqnarray}}
\newcommand{\eeqa}{\end{eqnarray}}
\newcommand{\RN}[1]{%
  \textup{\uppercase\expandafter{\romannumeral#1}}%
}

\graphicspath{{../figures/}}

\DeclareSIUnit\intensity{\watt\per\centi\meter\squared}
\DeclareSIUnit\fieldstrength{\volt\per\centi\meter}
\usepackage{xspace}

\DeclareUnicodeCharacter{2009}{\,}

\newcommand{\ie}{i.\,e.~}%

\newcommand{\Jvec}{$\boldsymbol{\mathrm{J}}$}
\newcommand{\Lvec}{$\boldsymbol{\mathrm{L}}$}

\newlength{\figwidth}
\newlength{\figwidthwide}
\let\orgautoref\autoref
\providecommand{\Autoref}{%
   \def\equationautorefname~##1\null{Equation~(##1)\null}%
  \def\figureautorefname{Figure}%
  \def\subfigureautorefname{Figure}%
   \def\sectionautorefname{Section}%
  \orgautoref}
\renewcommand{\autoref}{%
   \def\equationautorefname~##1\null{Eq.~(##1)\null}%
  \def\figureautorefname{Fig.}%
  \def\subfigureautorefname{Fig.}%
  \def\sectionautorefname{section}%
  \orgautoref}

\usepackage{color}
\usepackage[normalem]{ulem}
\definecolor{darkgreen}{rgb}{0.0,0.7,0.0}

\begin{document}

\title{A simple model for high rotational excitations of  molecules in a superfluid}

\author{Igor N. Cherepanov}
\affiliation{Institute of Science and Technology Austria, Am Campus 1, 3400 Klosterneuburg, Austria}

\author{Giacomo Bighin}
\affiliation{Institute of Science and Technology Austria, Am Campus 1, 3400 Klosterneuburg, Austria}
\affiliation{Institut f\"ur Theoretische Physik, Universit\"at Heidelberg, Philosophenweg 19, D-69120 Heidelberg, Germany}

\author{Constant A. Schouder}
\affiliation{Department of Chemistry, Aarhus University, 8000 Aarhus C, Denmark}

\author{Adam S. Chatterley}
\affiliation{Department of Chemistry, Aarhus University, 8000 Aarhus C, Denmark}




\author{Henrik Stapelfeldt}  
\email[Corresponding author: ]{henriks@chem.au.dk}
\affiliation{Department of Chemistry, Aarhus University, 8000 Aarhus C, Denmark}

\author{Mikhail Lemeshko}
\email[Corresponding author: ]{mikhail.lemeshko@ist.ac.at}
\affiliation{Institute of Science and Technology Austria, Am Campus 1, 3400 Klosterneuburg, Austria}

\date{\today}

\begin{abstract}
We present a simple quantum mechanical model describing excited rotational states of molecules in superfluid helium nanodroplets, as recently studied in  non-adiabatic molecular alignment experiments [Cherepanov \textit{et al.}, Phys. Rev. A \textbf{104}, L061303 (2021)]. We show that a linear molecule immersed in a superfluid can be seen as an effective symmetric top, similar to the rotational structure of radicals, such as OH  or NO, but with the angular momentum of the superfluid playing the role of the electronic angular momentum in free molecules. The model allows to evaluate the effective rotational and centrifugal distortion constants for a broad range of species and to explain the crossover between  light and heavy molecules in superfluid $^4$He in terms of the many-body wavefunction structure. Most important, the simple theory allows to answer the question as to what happens when the rotational angular momentum of the molecule increases beyond the lowest excited states accessible by infrared spectroscopy. Some of the above mentioned insights can be acquired by analyzing a simple $2 \times 2$ matrix.

\end{abstract}

\maketitle

\section{Introduction}
\label{sec:intro}

Interactions of individual molecules with superfluid helium-4 has been extensively studied during the last decades both experimentally and theoretically~\cite{ToenniesAngChem04, AncilottoIRPC17, VermaAdvPhys19}. According to infrared spectroscopy, the rotational motion of most molecules is affected by superfluid helium only quantitatively: while no drastic qualitative changes in rotational spectra is observed, the spectroscopic constants of molecules become ``renormalized'' due to the molecule--solvent interactions. In particular, the rotational constant, $B$, and the centrifugal distortion constant, $D$, assume different values as compared to gas phase molecules, $B^\ast < B$ and $D^\ast > D$. However, for the lowest $J$-levels, the rotational energy, $E_J$,  can still be accurately described by the gas-phase expression~\cite{GrebenevOCS,Nauta2001,nauta_vibrational_2001}:
\begin{equation}
E_J= B^* J(J+1) - D^*J^2(J+1)^2.
\label{eq:B*D*}
\end{equation}
Although there is little doubt that \autoref{eq:B*D*} describes the low-energy rotational structure for most molecules in superfluid $^4$He, little is known about the higher excited rotational states. In particular, we are talking about the states that are not not initially thermally populated due to the helium environment ($T\approx 0.37$~K in helium nanodroplets). Due to the spectroscopic selection rules, $\Delta J = \pm 1$,  conventional infrared and microwave spectroscopies are able to reach as far as only one rotational state above the initial Boltzmann distribution.

Theoretically, most quantum approaches focus on properties of molecules in superfluid-helium in the ground and the lowest excited rotational states~\cite{Hartmann1995, Lee1999, Kwon1999, CallegariPRL99, GrebenevOCS, LehmannJCP01, LehmannJCP02, Zillich2004, Zillich:2004cta}. The extension of \textit{ab initio} treatments to highly excited states, however, seems quite challenging \cite{Zillich2005}.

Recently it became possible to experimentally probe highly excited rotational states of molecules in helium nanodroplets using non-adiabatic alignment protocols~\cite{PentlehnerPRL13, Shepperson:2017gb, chatterley_rotational_2020, CherepanovPRA21}. Namely, analysing the Fourier transforms of alignment traces allowed to reveal the energies and lifetimes of rotational levels in superfluid $^4$He, up to $J \sim 16$. Moreover, the technique is applicable to molecules, that are non-responsive to infrared spectroscopy, such as I$_2$ and CS$_2$.

The goal of this paper is to present a simple quantum mechanical model that can be used to describe and to understand  rotational properties of molecules in a superfluid, including highly excited rotational states. Since such a many-body problem is extremely challenging to solve from first principles, we resort to a phenomenological treatment, based on the previously discussed  angulon model~\cite{SchmidtLem15, LemeshkoDroplets16, Lemeshko_2016_book}, which we simplify further in order to make it more transparent.

The present study builds upon our recent experimental and theoretical work~\cite{CherepanovPRA21}. However, apart from describing the theoretical machinery of our model in more detail, this paper provides several novel  insights, such as comparisons of spectroscopic constants for a broad range of molecular species and explaining the origin of the crossover between the light and heavy molecules well known in helium droplet spectroscopy~\cite{ToenniesAngChem04}.   The main theoretical message of Ref.~\cite{CherepanovPRA21} was, on the other hand, the possibility to describe the rotational spectrum in terms of the angular momentum transfer between the molecule and the superfluid. Therefore, here we are going to omit a detailed discussion of the angular momentum coupling and refer the interested reader to our earlier work. 

We start by describing the model in \Autoref{secH} and show that a linear molecule in superfluid helium can be seen as an effective symmetric top. This description is similar to that of open-shell molecules like OH or NO~\cite{LevebvreBrionField2, BrownRot}, but with the angular momentum of the superfluid playing the role of the angular momentum of the electronic shell.  In \Autoref{sec:excited} we analyze the energy level structure of such an effective symmetric top and gather insights relevant to experiments on molecules in He droplets. In \Autoref{sec:2level} we show how the model can be simplified even further and that important insights can be gathered from solutions of a $2\times 2$ matrix. Finally, in  \Autoref{sec:BandD} we present the results for effective spectroscopic constants, $B^*$ and $D^*$, and compare them with experiment.  \Autoref{sec:conclusions} provides the conclusions of this study.

\section{A Solvated linear Molecule  \\[3pt]  becomes  a symmetric top}
\label{secH}

\subsection{The angulon Hamiltonian}
\label{sec:ham}

We consider a linear molecule with a rotational constant $B$ revolving in the bath of bosons (collective excitations in $^4$He). To obtain the simplest possible model, we take into into account only a single mode of the bath with energy $\omega$ and angular momentum $\lambda$. In the case of superfluid
helium it might be tempting to label these excitations as rotons, however we intentionally keep the treatment as general as possible. In addition, we take into account only the linear molecule--He coupling term. This corresponds to a further simplification of the previously developed angulon model~\cite{SchmidtLem15, LemeshkoDroplets16}. In the molecular (body-fixed) frame, the system is described by the following Hamiltonian \cite{SchmidtLem16}:

\begin{equation}
\label{eq:Hamiltonian}
\hat{H}=B(\boldsymbol{\mathrm{\hat{L}}}-\boldsymbol{\mathrm{\hat{\Lambda}}})^2 +  \omega \sum_{n} \hat{b}^{\dagger}_{\lambda n}\hat{b}_{\lambda n}+ u \big(\hat{b}^{\dagger}_{\lambda 0} +\hat{b}_{\lambda 0}\big) \;
\end{equation}
where $\hat{b}^{\dagger}_{\lambda n}$ ($\hat{b}_{\lambda n}$) create (annihilate) a bosonic excitation with angular momentum $\lambda$ and projection onto the molecular (i.e.\ interatomic) $z$-axis $n$, $u$ reflects the strength of the anisotropic molecule-bath interaction.   $\boldsymbol{\mathrm{\hat{L}}}$ is the total angular momentum of the system and $\boldsymbol{\mathrm{\hat{\Lambda}}} = \sum_{n \nu} \hat{b}^{\dagger}_{\lambda n} \boldsymbol{\mathrm{\sigma}}^{\lambda}_{n \nu} \hat{b}_{\lambda \nu}$ defines the angular momentum acquired by the bath. Here, $\boldsymbol{\mathrm{\sigma}}^{\lambda}_{n \nu}$ denotes the angular momentum matrices fulfilling the $SO(3)$ algebra in the representation of angular momentum $\lambda$.

In this paper we focus on the weak-coupling theory, that is, we start from a non-interacting case, corresponding to no helium excitations and add excitations one by one. The weak coupling angulon theory accounting for a single excitation of helium was shown to predict renormalization of rotational constants of light molecules trapped inside helium nanodroplets in good agreement with experimental data \cite{LemeshkoDroplets16}. To accurately describe heavy rotors, one has to deal with more sophisticated solutions of the Hamiltonian, \autoref{eq:Hamiltonian}. They involve perturbations on the top of a microscopic deformation of the helium bath, \ie an infinite number of bosonic excitations \cite{SchmidtLem16, Bighin17, Bighin18}. In the course of the paper, however, we aim to demonstrate that the solutions including up to triple excitations are able to catch changes in molecular spectra for broad range of species measured in helium.

The first term of~\autoref{eq:Hamiltonian} represents an effective symmetric-top Hamiltonian, similar to that used to describe the electronic states of radicals, such as NO or OH~\cite{LevebvreBrionField2, BrownRot}. In our case, the boson angular momentum $\boldsymbol{\mathrm{\hat{\Lambda}}}$ plays the role of the electronic angular momentum in open-shell molecules. The corresponding rotational states can be expressed through the symmetric-top states $\ket{L N M}$, where $N$ and $M$ label the projections of the \textit{total} angular momentum, $\mathbf{L}$, on the molecular and space-fixed axes, respectively. For a linear molecule, the projection of the molecular rotational angular momentum, $\mathbf{J}$, on the molecular $z$-axis is zero, therefore $N$ entirely corresponds to the projection of $\mathbf{\Lambda}$. In other words, the interaction with the superfluid, $u$ of~\autoref{eq:Hamiltonian}, creates some non-zero angular momentum $\mathbf{\Lambda}$ that can be seen as analogous to the electronic angular momentum of open-shell molecules. Or, semiclassically speaking,  a ``nonsuperfluid shell" of He atoms attached to the linear molecule, provides it with an additional ``thickness'', hence the symmetric-top description. The classification of different $\mathbf{L} - \mathbf{\Lambda}$ coupling schemes in terms of Hund's cases (in analogy with gas-phase species) is another interesting problem that is not going to be discussed here. Furthermore, we omit the detailed discussion of molecule-bath angular momentum transfer, that has already been presented elsewhere~\cite{CherepanovPRA21}.

\subsection{Basis states and diagonalization}

It is worth noting that in the case of a particle linearly moving in a bosonic environment (the so-called ``polaron problem''), writing the Hamiltonian in the frame co-moving with the particle (by analogy with \autoref{eq:Hamiltonian})  allows to completely decouple the particle and environment degrees of freedom~\cite{Devreese15}. This is impossible to do for the case of a rotating molecular impurity, since different components of the angular momentum $\mathbf{\hat{L}}$ do not commute with each other and it is therefore impossible to replace $\mathbf{\hat{L}}$ in~\autoref{eq:Hamiltonian} by a classical number $L$, as one could do for the total linear momentum operator, $\mathbf{\hat{P}} \to P$. Although the magnitude of the total angular momentum, $L$, is conserved, a general solution is going to be a superposition of states corresponding to different projections $N$, which, in turn, can contain different numbers of bosonic excitations (the $M$-quantum number plays no role in the absence of external fields).

We diagonolize the Hamiltonian, \autoref{eq:Hamiltonian}, in the following basis:
\begin{equation}
\psi_{L [n_1 n_2...n_m], M}^{(m)} =  \ket{LNM}_\text{mol} \left( b^{\dagger}_{\lambda n_1} b^{\dagger}_{\lambda n_2} ...  b^{\dagger}_{\lambda n_m}\ket{0}_\text{bos} \right)
\label{eq:psi}
\end{equation}
$N=\sum_{i=1}^{m} n_i$ and $M$ refer to the total projection of $\boldsymbol{\mathrm{\hat{L}}}$ on the molecular and laboratory $z$-axis, respectively. $N$ and $M$ take values in the range $[-L,L]$. An additional condition on the total projection $N$ is imposed by the following limitation on $n_i$: $\vert n_i \vert \leq \lambda$. As stated above, we restrict our basis set to $m \leq 3$. Note that although \autoref{eq:Hamiltonian} is a substantial simplification of the original angulon Hamiltonian, the ansatz of \autoref{eq:psi} represents a substantially expanded basis set compared to the previous treatments, where only single excitations ($m=1$) were taken into account~\cite{SchmidtLem15}. Including multiple bath excitations allows to describe a broader range of molecules using the weak-coupling theory.

The $m=0$ case describes a bare (``gas phase'') molecular state $\psi^{(m=0)}_{LM} = \vert L, N=0, M \rangle_{\text{mol}}\ket{0}_\text{bos}$. Diagonalization of the Hamiltonian in this basis obviously leads to the $(2L+1)$-fold degenerate energy spectrum of an isolated rigid rotor,  $BL(L+1)$. The projection $N$ equals to zero in the absence of the excitation since we assume that the molecule is linear. The $m>0$ cases introduce multiple excitations of the bath $b^{\dagger}_{\lambda n_1} b^{\dagger}_{\lambda n_2} ...  b^{\dagger}_{\lambda n_m}\ket{0}_\text{bos}$.

In our model, the molecule can directly induce only deformations of the boson density that preserve $N=0$. Thus, they are strongly aligned along the molecular $z$-axis. This can be seen from the third term in \autoref{eq:Hamiltonian} and from the corresponding density plot for $L=0$ in \autoref{fig:band}(a). Nevertheless, the presence of the spin-orbit--like (or Coriolis-like)  interaction, the  $- 2 \boldsymbol{\mathrm{\hat{L}}} \cdot \boldsymbol{\mathrm{\hat{\Lambda}}}$ term in  \autoref{eq:Hamiltonian}, causes precession of $\boldsymbol{\mathrm{\hat{\Lambda}}}$ about the molecular $z$-axis, somewhat similar to a spin in a magnetic field.  Minimization of the angle between $\boldsymbol{\mathrm{\hat{L}}}$ and $\boldsymbol{\mathrm{\hat{\Lambda}}}$ (which, in turn, minimises the energy of the system) leads to increase in $N$ and hence to the wider distribution of the bosons density with respect to the molecular $z$-axis as shown in \autoref{fig:band}(a). As a result, the linear molecule dressed by a cloud of excitations resembles a symmetric top whose non-zero projection $N$ is exclusively provided by the angular momentum of the He atoms in the solvation shell. In the following sections we discuss how the spectrum of such an effective symmetric top differs from the quadratic spectrum of a rigid linear rotor.  

\section{Excited rotational states  \\[3pt] in the superfluid}
\label{sec:excited}

We start from exploring the stationary states of the system, previously briefly described in Ref.~\cite{CherepanovPRA21} In what follows, we show that through analysing the states of an effective symmetric top (cf.\ \Autoref{sec:ham}), one can understand the distribution of angular momentum due to the molecule--helium interaction and how it changes in an external laser field.

\begin{figure}[t]
\centering
\includegraphics[width=0.9\linewidth]{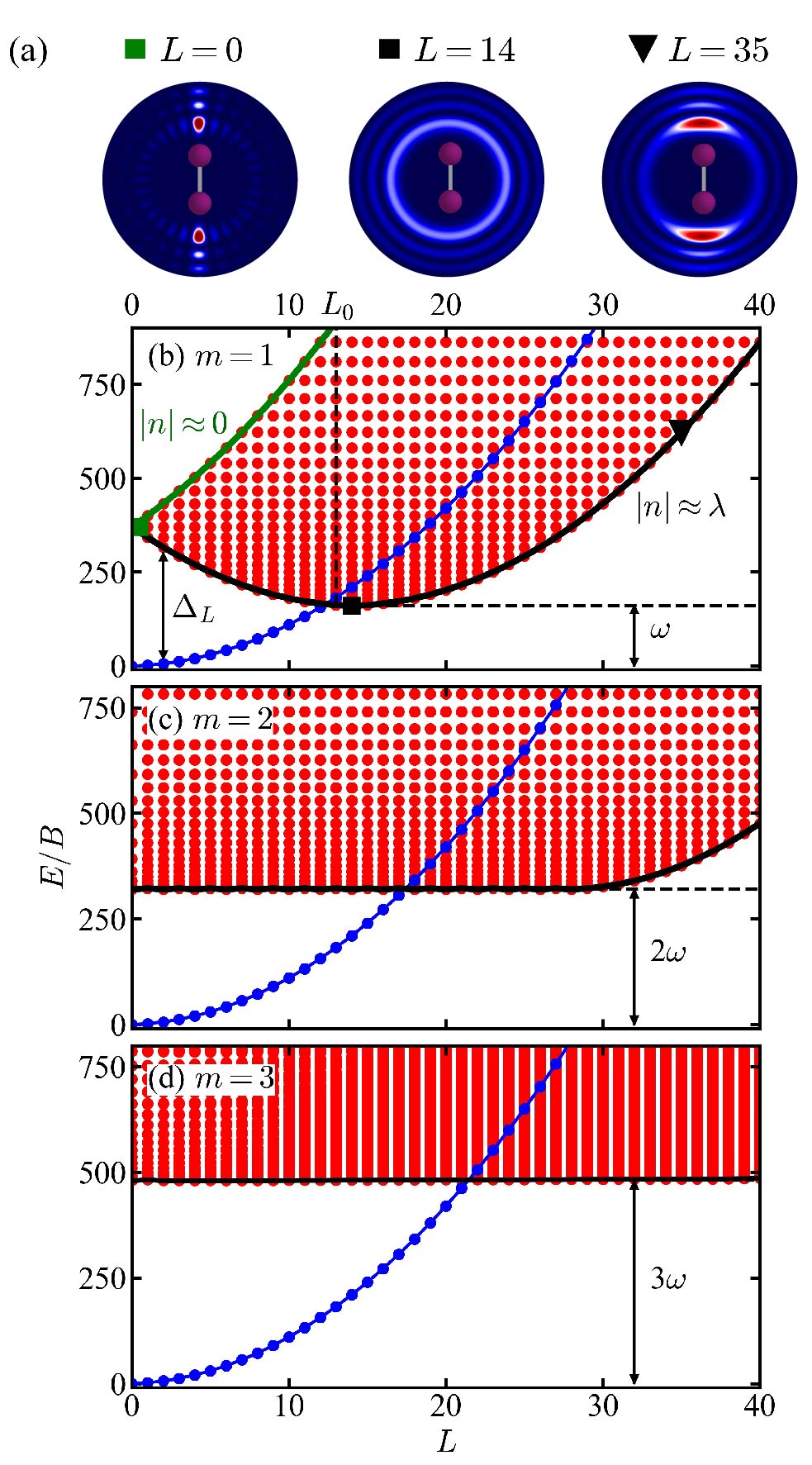}
\caption{(a) Boson density distributions  in the molecular (body-fixed) frame for selected excited states (marked by the corresponding symbols in (b)). (b)--(d) Energy diagram of the band of excited states involving single ($m=1$), double ($m=2$), and triple ($m=3$) excitations of the bosonic bath (red dots), respectively. The gas-phase rotational spectrum ($m=0$) is shown by the blue dots. The green line indicates configurations with projection $n \approx 0$ onto molecular $z$-axis, the black lines denote the excited states with minimum energy and largest possible total projection, $N=\sum n_i$, for a given $L$. }
\label{fig:band}
\end{figure}

\Autoref{fig:band} shows the possible states of the system for the case of one, two and three bath excitations.  Each dot in \autoref{fig:band}(b--d) represents a unique configuration, their energies are obtained by diagonalization of the Hamiltonian, \autoref{eq:Hamiltonian}, assuming $u=0$. To facilitate the visualisation of each contribution,  we perform diagonalization in each of three bases, \autoref{eq:psi} with $m = 1-3$, separately and plot the calculated energies in \autoref{fig:band}(b--d), respectively. The blue dots correspond to $m=0$, \ie to the energies of an isolated gas phase molecule, $BL(L+1)$. In this case, the molecular angular momentum equals to the total angular momentum \Jvec~$=$ \Lvec, no excitations of the bath are present.

The red dots in \autoref{fig:band}(b--d) form the band of excited states. For illustrative purposes, \autoref{fig:band}(a) shows the molecular-frame densities  of He corresponding to three of these excited states at $L=0, 14$ and $35$, also marked in \autoref{fig:band}(b). In these configurations, the total angular momentum \Lvec~is shared between the molecule and the helium excitations. For simplicity we begin with the states involving single excitations, $m=1$, carrying energy $\omega$ and angular momentum $\lambda$ with projection $n$ onto the molecular $z$-axis, as shown in \autoref{fig:band}(b).  Neglecting the off-diagonal  $\mathbf{\hat{L}}_\pm \mathbf{\hat{\Lambda}}_\mp$ terms in \autoref{eq:Hamiltonian}, the energies in the  $\ket{L n M}$-basis are given by: 
\begin{equation}
\label{eq:oblate}
E^{\lambda}_{L,n}= BL(L+1) -2Bn^2 +B\lambda(\lambda+1) + \omega,
\end{equation}
where we introduced an additional shift by the excitation energy $\omega$. \autoref{eq:oblate} corresponds to the energies of an oblate (disk-shaped) symmetric top, shifted by $B\lambda(\lambda+1) + \omega$ from zero. Since the off-diagonal components of the $- 2 \boldsymbol{\mathrm{\hat{L}}} \cdot \boldsymbol{\mathrm{\hat{\Lambda}}}$ term in \autoref{eq:Hamiltonian} mix $n$, the resulting state in the most general case corresponds to a superposition of different $n$ projections. From the shape of the band of excited states in \autoref{fig:band}(b) one can see that the energetics remains similar to that of an oblate symmetric top even when the off-diagonal terms are fully taken into account. In particular, the system tends to occupy the states with non-zero $n$.

To provide an intuitive understanding of the perturbations caused by molecular rotation, we plot the distribution of helium density in the molecular frame for selected excited states in \autoref{fig:band}(a). Note that these densities are obtained at $u=0$ and do not correspond to the density deformations induced by the molecule (as discussed in Ref.~\cite{CherepanovPRA21}). Instead, these plots are supposed to illustrate how the excited bath states look like in real space in the absence of molecule--helium interactions.


Let us consider a particular excited state at $L=0$ with a well-defined projection, $n=0$, marked by the green square in \autoref{fig:band}(a--b). The energy cost to create such an excitation is $\omega+B\lambda(\lambda+1)$. The angular density distribution plotted in \autoref{fig:band}(a) shows that the bosons primarily reside at the poles of the molecule (linear configuration). As $L$ increases, the states with the dominating zero projection contribution form the upper edge of the band in \autoref{fig:band}(b) coloured in green. Classically, they might be thought of as rigid rotation of the molecule with its solvation shell.

However, the states with $n \approx 0$ are not the ground state of an effective oblate top described by \autoref{eq:oblate}. For $L>0$, as soon as the excitation is created, the Coriolis coupling $-2\boldsymbol{\mathrm{\hat{L}}}\cdot \boldsymbol{\mathrm{\hat{\Lambda}}}$ makes the $n \neq 0$ configurations energetically more favorable. The bosons density shifts to the waist of the molecule ($T$-shape configuration) as $L$ increases. The states with the maximum $\vert n \vert $ build the lower edge of the band in \autoref{fig:band}(b) coloured in black. Its parabolic shape is defined by the above mentioned restrictions set on $n$: (i) $\vert n \vert \leq L$ and (ii)  $\vert n \vert \leq \lambda$. The minimum energy equals to $\omega$ and it is reached at $L=\lambda$. Furthermore, the lowest excited state at $L=\lambda$ shows a perfectly uniform distribution over $n$. This state is marked by the black square in \autoref{fig:band}(a--b). Its angular density distribution is delocalized as plotted in \autoref{fig:band}(a). In the classical picture, we interpret these observations as manifestation of non-rigidity of the molecule--bosons coupled rotation. 

Further growth of $L$ beyond $\lambda$, nevertheless, leads to the bending up of the lower edge of the band since $\vert n \vert$ can no longer increase. In \autoref{fig:band}(a--b) we mark one of the states satisfying $L \ll \lambda$ by the black triangle. The boson density moves back towards the poles of the molecule. We would like to stress that these findings are analogous to the resonance behaviour of the helium anisotropy found within the semiclassical toy model \cite{LehmannJCP01}. In that model, the solvation shell is modelled as a ring of $N_{\text{He}}$ helium atoms. Identically to $\lambda$ in our model, $N_{\text{He}}$ determines the symmetry of the helium solvation shell. The maximum anisotropy observed at $L=N_{\text{He}}$ draws parallels to the results discussed above.

Qualitatively, similar considerations are valid for double ($m=2$) and triple ($m=3$) excitations. The corresponding energy diagrams are shown in \autoref{fig:band}(c--d). The only noticeable difference arises from the possibility to sum up individual projections $n_i$ to the total projection $N$. The constraint $\vert n \vert \leq L$ is thereby lifted which substantially expands the Hilbert space of the bath excitations. As a consequence, the lower edge of the band in the range of $L<\lambda$ becomes flat. It happens due to the fact that the combinations of several excitations having the largest possible projections $\vert n \vert = \lambda$ of the opposite sign are allowed even for small $L$. The minimum energy therefore reduces to $m \omega$, the lower edge of the band starts bending upwards at $L = m \lambda$.


\section{Even simpler: a two-level model}
\label{sec:2level}

In the previous section we discussed the possible states of the ``many-body symmetric top'' without explicitly taking into account the coupling between these states induced by the molecule--helium interactions. A non-zero value of $u$ results in coupling of the bare molecular state ($m=0$) to the excited states with $m>0$ discussed above. The deviations of the final energies with respect to the gas-phase spectrum describe the net effect of the surrounding environment on molecular rotation. These perturbations can be detected in experiments as a change of the effective spectroscopic constants and are therefore of particular interest. While it is possible to evaluate them numerically, we would like to focus on the aspects of the model available for analytical treatment at first.

In our model, bare rotational states couple in  first order only to the single excitations with $n=0$ (cf. the third term in \autoref{eq:Hamiltonian}). For small $L$, the gas-phase energies and the band of excited states are separated by the relatively large energy gap, $\Delta_0$, as compared to the rotational kinetic energy:
\begin{equation}
\Delta_0 = \Delta_{L=0} = \omega+B\lambda(\lambda+1).
\label{eq:delta0}
\end{equation}
Note that the gap depends on $B$ and never closes for small $L$. In particular, this means that a few well-distinguished rotational levels must be present even for very light rotors with $B$ exceeding $\omega$, as confirmed by  experiment~\cite{ToenniesAngChem04}.  Since it is hard to obtain an accurate analytical expression for the gap, $\Delta_L$, for an arbitrary $L$, we use its numerically calculated values shown in \autoref{fig:band}(b). Nevertheless, one can say that in the linear approximation its slope is approximately given by $B\lambda$. The ratio $u/\Delta_L$ and its dependence on $L$ define how strong the bath perturbs the molecular energies. If the interaction strength is comparable to or exceeds the kinetic energy of the excitation, $u \gtrsim \Delta_L$, the rotational spectrum is subject to strong renormalization. In the opposite case of $u \ll \Delta_L$, the molecule does not experience a strong influence from the bath.

\begin{figure}[t]
\centering
\includegraphics[width=\linewidth]{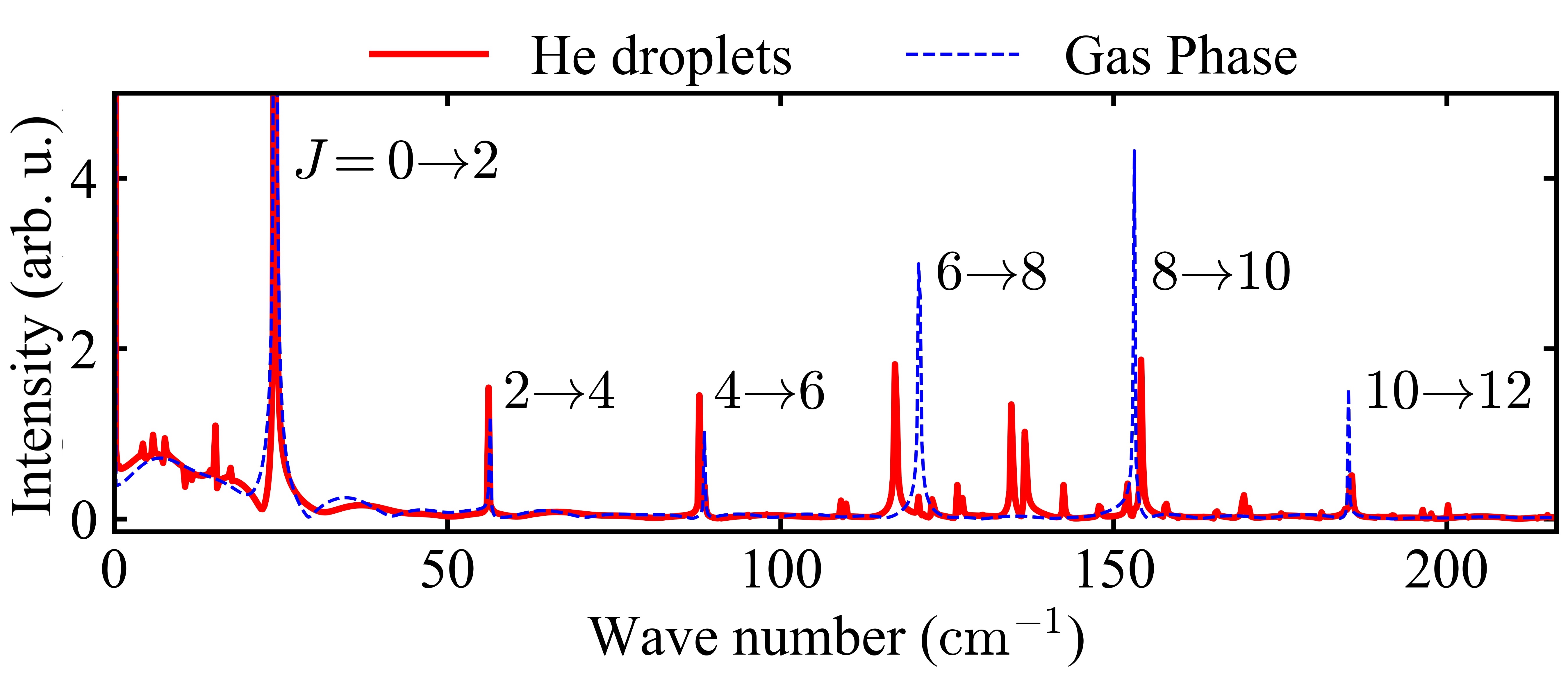}
\caption{Raman-like ($\Delta L = \pm 2$) rotational spectra for a typical light rotor molecule ($B = 4~\ce{cm^{-1}}$) in helium ($u = 10~\ce{cm^{-1}}$, red solid line) and in the gas phase (blue dashed line). The molecule initially resides in the ground $L=0$ state.}
\label{fig:splitting}
\end{figure}

In \autoref{fig:splitting} we calculate the Raman--like ($\Delta L = \pm 2$) rotational spectrum for a typical light molecule with $B = 4~\ce{cm^{-1}}$ and $u/\Delta_L \approx 0.05$. Although the energy of the first excited state, $L=1$, exceeds $\omega$, we see three well-defined spectral lines attributed to the total angular momentum states up to $L=6$. Because of the small $u/\Delta_L$ ratio, they exhibit a slight red shift and minor changes in intensity distribution in comparison to the gas phase. This coincides with a few percent change in the rotational constants observed for light molecules~\cite{ToenniesAngChem04}. A secondary substructure originating from the perturbed band of the excited states is separated from the main peak by the energy gap of order $\Delta_L$. We do not label these spectral features in the figure because of their negligible spectral weight.

The situation changes radically  when $L$ is further increased. As may be seen in \autoref{fig:band}(b), the $L$-dependent energy gap $\Delta_L$ shrinks and finally closes at
\begin{equation}
L_0 \approx \left( \frac{\omega+B\lambda(\lambda+1)}{2B} \right)^{1/2}.
\label{eq:L0}
\end{equation}
Referring back to \autoref{fig:splitting}, we observe that the lines involving the states $L \gtrsim L_0$ ($L_0 = 8$ in this particular case) develop a rich substructure consisting of multiple secondary peaks. In principle, all of them might be ascribed to the transitions between excited states that preserve $N$ but change $L$ according to the selection rules $\Delta L = \pm 2$. We expect that these lines will be substantially broadened if one goes beyond the single mode approximation and includes a full continuous dispersion of bulk helium $\omega (k)$ into the model. This effect has been demonstrated in Ref. \cite{Cherepanov17} for symmetric top molecules. It quantitatively explains the anomalous broadening of spectral lines, initially observed in experiments with \ce{CH3} \cite{Morrison_2013} and \ce{NH3} \cite{Slipchenko2005} in helium droplets.

Next, we derive simple analytical formulas for the renormalized spectroscopic constants. In first order, the gas-phase rotational states are coupled only to the states with $n = N = 0$ and $m=1$, which are, in turn, coupled to states with nonzero $N$ and $m \geq 1$  in higher orders. To simplify the problem, we can assume that all higher-order interactions can be incorporated into an effective energy shift,  $\delta_L$, of the single excitations with $N=0$ with respect to the energy given by \autoref{eq:oblate}, $E^{\lambda}_{L,0} = \Delta_0+BL(L+1)$. 

In such a way, we can qualitatively describe the  $L$-dependent deformations of the gas phase spectrum as coupling of the bare molecular states to a single ``dressed'' $N=0$, $m=1$ state for a given $L$, which corresponds to an effective two-level system:
\begin{equation}
\hat{H}'_L = \begin{bmatrix} BL(L+1)&u\\u&BL(L+1)+\Delta_0-\delta_L \end{bmatrix}.
\label{eq:tls}
\end{equation}
 After dropping $L$-independent contributions, the ground state energy of the Hamiltonian \eqref{eq:tls} reads:
\begin{equation}
E_L = BL(L+1) -\frac{\delta_L}{2} - \frac{\sqrt{(\Delta_0-\delta_L)^2 + 4u^2}}{2}.
\label{eq:tls_gs}
\end{equation}
Based on \autoref{eq:oblate}, we set $\delta_L= 2B \gamma L(L+1)$ with the parameter $\gamma \in [0,1]$ defining how strong is the effect of high order interactions on the $N=0$, $m=1$ states, i.e. how much their energy effectively shifts from $E^{\lambda}_{L,0}$ (located close to the green line in \autoref{fig:band}(b)) towards the lower edge of the band (black line in the same figure).  In both limits of light ($B \to \infty$) and heavy ($B \to 0$)  rotors,  $E_L$ can be expanded in a series:
\begin{equation}
E_L = B^*L(L+1) - D^*L^2(L+1)^2 + O\bigg(L^3(L+1)^3\bigg),
\label{eq:expanision}
\end{equation}
cf.~\autoref{eq:B*D*}. For light rotors (LR), we make use of the condition $u \ll \Delta_0$ to show that the zero-order term in $u$ cancels out leading to weak renormalization of spectroscopic constants:
\begin{equation}
\frac{B^*_{\text{LR}}}{B}  \approx 1- \frac{2 \gamma u^2}{\Delta_0^2};~D^*_{\text{LR}} \approx \frac{4B^2 \gamma^2 u^2}{\Delta_0^3}
\label{eq:weak_constants}
\end{equation}
These expressions coincide with the exact analytical results obtained for small $L$ in Ref. \cite{CherepanovPRA21}. Both renormalized spectroscopic constants contain the small parameter $u/\Delta_0$ which guarantees that  $B^* \to B$ and $D^* \to 0$ in the free-rotor limit.

In the opposite limit of heavy rotors (HR), the expansion of energy in powers of a small parameter $(\frac{\Delta_0-\delta_L}{2u})^2$ gives
\begin{equation}
\frac{B^*_{\text{HR}}}{B} \approx 1-\gamma;~D^*_{\text{HR}} \approx \frac{B^2 \gamma^2}{2u}
\label{eq:strong_constants}
\end{equation}
 Since the average value of the parameter is $\gamma \sim 1/2$, the rotational constant shows non-negligible renormalization in this case. The expression for $D^*_{\text{HR}}$ closely resembles the empirical formula $D^* = 0.031 \times B^{* 1.818}$ found in Ref.~\cite{Choi2006} by fitting to the experimental data (setting $\gamma = 1/2$ and $u=10$ gives the prefactor of $\approx 0.01$). Furthermore, \autoref{eq:strong_constants} predicts the same dependence on $B$ as the approximate solutions of the strong coupling model reported in Ref. \cite{LemeshkoDroplets16}.


\section{Effective spectroscopic\\[2pt] constants}
\label{sec:BandD}


Over the past two decades a lot of experimental and theoretical data were collected for effective spectroscopic constants of a broad range of molecular species in superfluid helium (see e.g. Ref.~\cite{ToenniesAngChem04}). Although the main focus of this paper is on highly excited rotational states, benchmarking the qualitative results of the theory against the available experimental data is a good test of the model. In this Section, we work with the numerical solutions of the full model developed in \Autoref{secH}, as opposed to the simplified solutions discussed in the previous section. The values of $B^*$ and $D^*$ discussed below were obtained by fitting the energies of the $L=0-3$ states to \autoref{eq:B*D*}.

\begin{figure}[t]
\centering
\includegraphics[width=\linewidth]{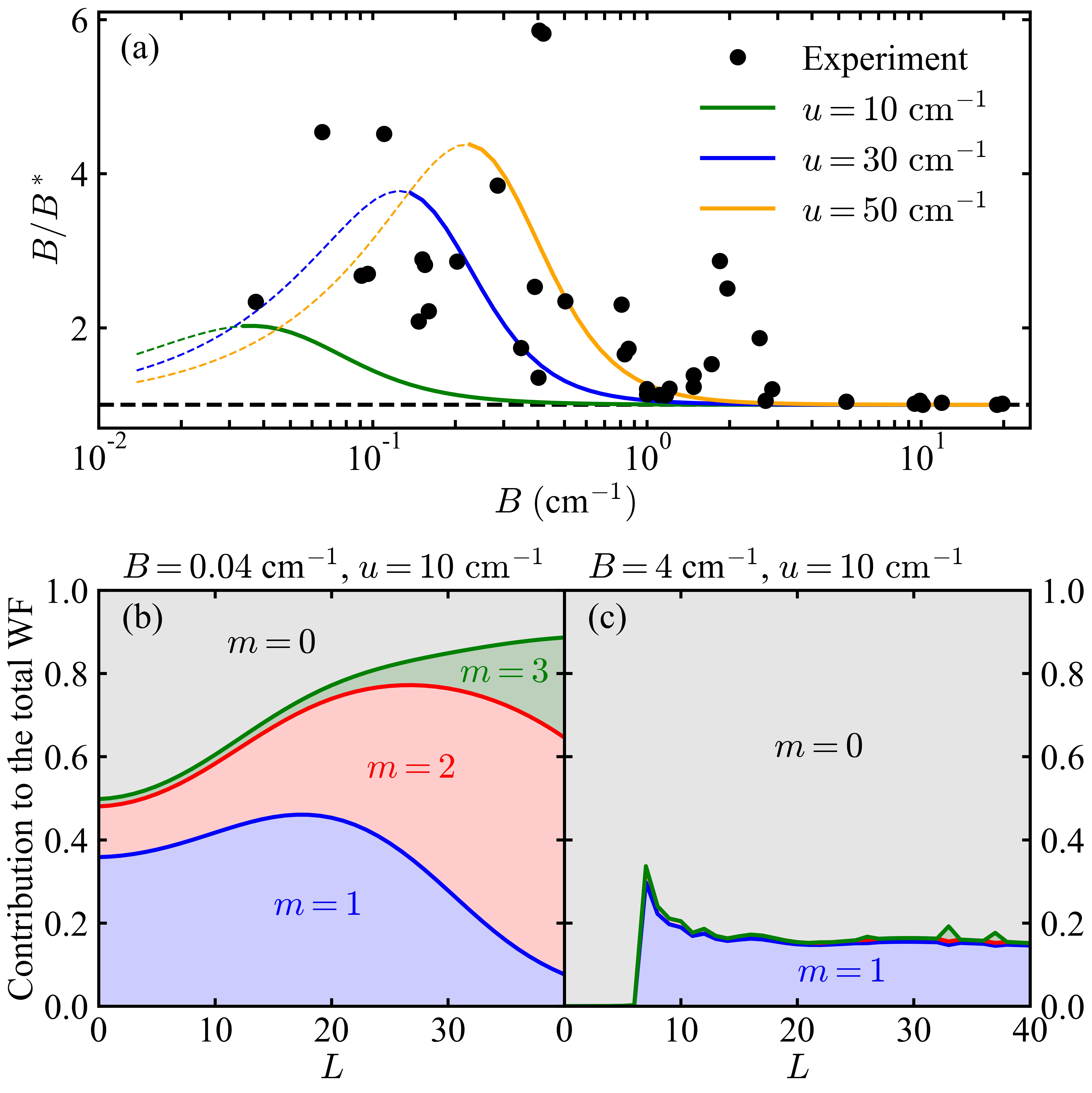}
\caption{(a) The reduction factor of rotational constants, $B/B^*$, for molecules in helium as a function of the gas phase rotational constant, $B$. The solid lines show the theoretical predictions obtained in the present work for selected values of the interaction strength, $u$. The dotted part of the lines indicates the range of $B$ where the weak coupling theory fails. Experimental data points are shown by the black circles. (b),(c) The relative contributions of the states involving single ($m=1$), double ($m=2$), triple ($m=3$) excitations of helium as well as bare molecular states ($m=0$) to the total wave function for a heavy rotor ($B = 0.04~\ce{cm^{-1}}$) and a light rotor ($B = 4~\ce{cm^{-1}}$), respectively. The interaction parameter $u = 10$~cm$^{-1}$ in both cases. }
\label{fig:renorm}
\end{figure}

In \autoref{fig:renorm}(a) we compare the effective rotational constants ($B/B^*$ as a function of $B$) obtained within our model (lines) with the results of experiments (black circles) \cite{Hartmann1995, Harms:1997iv, Lee1999, Callegari_2000, Conjusteau_2000, Nauta_2000, GrebenevOCS, Callegari2001, Nauta2001, Nauta_2001, Madeja_2002, Poertner_2002, Zillich2004, ToenniesAngChem04, Lindsay_2005, Paesani:2005jr, Slipchenko2005, vHaeften05, Scheele_2005, Choi2006,  Kuyanov2006, Skvortsov_2007, Hoshina2010, Raston_2011, Raston_2012, Morrison_2013, Raston_2013, Raston_2014, Faulkner_2018, chatterley_rotational_2020, Raston_2021}. The energy of the bosonic mode $\omega$ was fixed to $\SI{6}{cm^{-1}}$, the roton energy of bulk helium \cite{Donnelly1981}; $\lambda$ was set to 14, this choice is motivated by the results of Ref. \cite{CherepanovPRA21}. The molecule--helium coupling constant, $u$, depends on the details of the molecule--He potential energy surface (PES) and is going to be different for each molecule. Moreover, $u$ does not show any significant correlation with $B$. According to Ref. \cite{LemeshkoDroplets16}, the interaction parameter extracted from the molecule--He PES and expressed in absolute units varies within one order of magnitude for the species whose rotational constant cover more than three orders of magnitude. Our goal is to focus on the general trend, therefore, we present the theoretical curves for three different values of $u$ and the experimental data on $B/B^*$, without discussing concrete molecular species.

The overall trend seen in \autoref{fig:renorm}(a) can be explained semiclassically by the ``adiabatic following" model \cite{Lee1999, PatelJCP03, Markovskiy_2009}, revealing the crossover between the heavy and light species. In a simple picture, heavy rotors ($B \lesssim \SI{1}{cm^{-1}}$) rotate slow enough for the helium solvation shell to follow. Such strong coupling leads to a significant reduction of $B$, up to a factor of 6. Light rotors ($B \gtrsim \SI{1}{cm^{-1}}$), in contrast, rotate so fast that they decouple from helium and their rotational constant is almost not renormalized.

Note that the ansatz of \autoref{eq:psi} corresponds to the weak-coupling approximation, which breaks down in the limit of $B \to 0$, i.e. for very heavy molecules. The results furnished by the model in this regime (dotted lines \autoref{fig:renorm}(a)) are unphysical. This behaviour might also be rationalized within the effective two-level model of \Autoref{sec:2level}. If $B \to 0$, the shift $\delta_L$ in \autoref{eq:tls_gs} vanishes, thereby eliminating the $L$-dependence from the model (or alternatively, the lower edge of the band in \autoref{fig:band}(b) becomes flat). Although the admixture of bosonic excitations into the total wave function might be dominant, it does not bring any $L$-dependent contribution to the energy. The decreasing renormalization in this region is, thus, of a completely different nature than in the case of light rotors. Including the excitations with $m >3$ into the basis may substantially improve solutions of the model Hamiltonian, \autoref{eq:Hamiltonian}), in this regime.
 
 \begin{figure}[t]
\centering
\includegraphics[width=\linewidth]{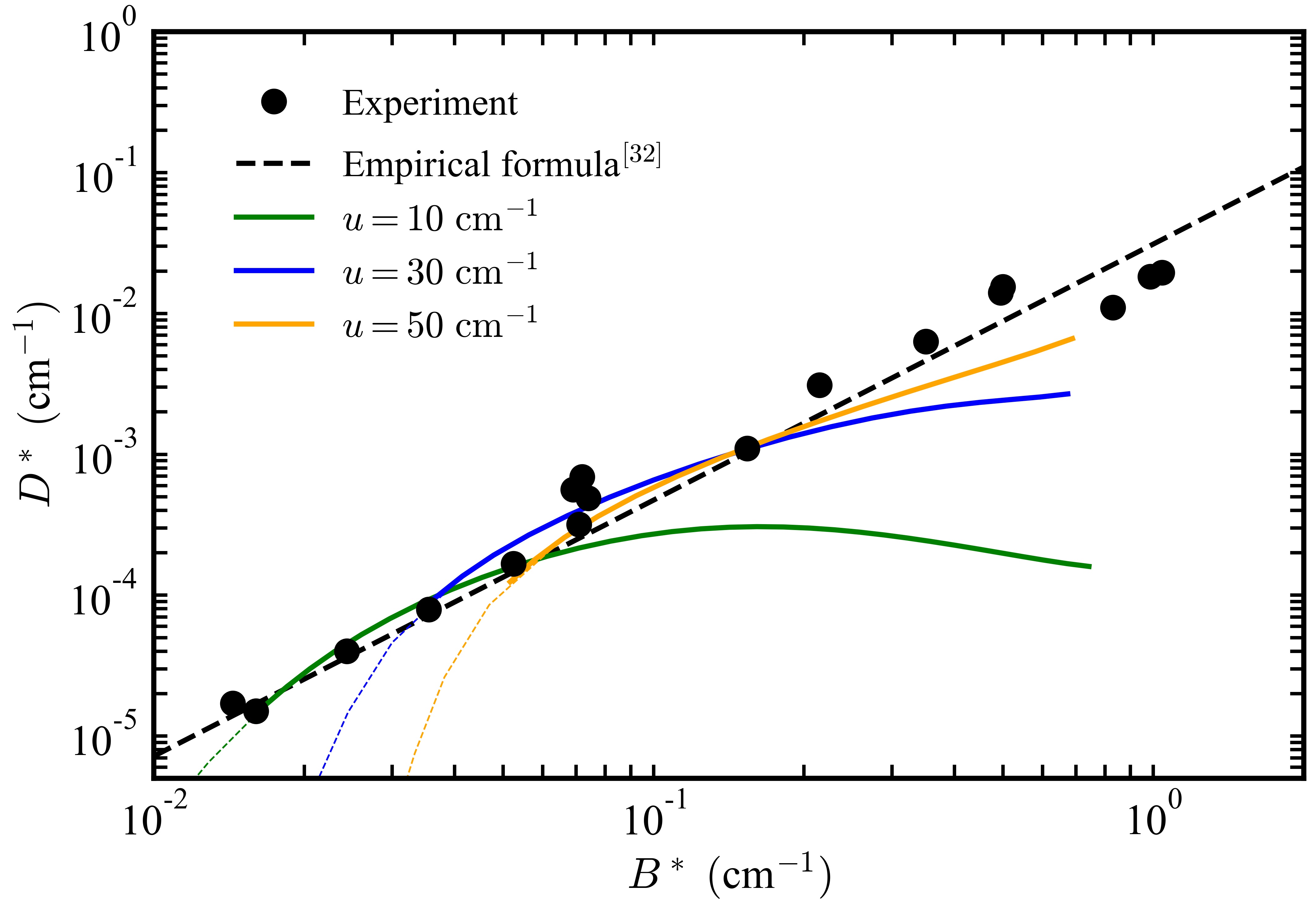}
\caption{The effective centrifugal distortion constant, $D^*$, as a function of the effective rotational constant $B^*$ (solid lines) for selected values of the interaction parameter $u$. The dotted parts of the lines indicate the range of $B^*$ where the weak-coupling theory fails. Experimental data points are shown by the black circles. The dashed black line corresponds to the empirical formula from Ref. \cite{Choi2006}}.
\label{fig:D}
\end{figure}

Figures~\ref{fig:renorm}(b) and (c) break down the contributions of different numbers of helium excitations into the total wavefunction for heavy and light molecules, respectively. The coupling parameter is set to the same value of $u = 10~\ce{cm^{-1}}$ in both cases. For heavy  molecules, $B = 0.04~\ce{cm^{-1}}$, \autoref{fig:renorm}(b), the contribution of the bare molecular state, $m=0$, is approximately 50\% for $L=0$ and monotonously decreases with $L$, while higher excitations $m=2$ and $3$ get more populated. For light molecules, on the other hand, there is a sharp transition point $L_0$, such that for $L < L_0$ only $m=0$ states are populated, while for $L>L_0$ also the states with nonzero $m$ are, see the example for $B = 4~\ce{cm^{-1}}$ in \autoref{fig:renorm}(c).

 This shows an important difference between heavy and light molecules, previously broadly discussed in the literature from other points of  view~\cite{ToenniesAngChem04}. For heavy molecules, even in the absence of rotation, the molecule--helium interaction distorts the surrounding superfluid and creates He excitations co-rotating with the molecule (a ``non-superfluid solvation shell''). For light molecules at small $L$ the bath excitations are only virtual (in agreement with the results of Ref.~\cite{LemeshkoDroplets16}), resulting in a very small $B$-renormalization. After some critical value of $L \sim L_0$, the bare molecular states cross the excitation threshold and start coupling to the bath strongly, which results in substantial population of $m\neq0$ states.

\Autoref{fig:D} shows the effective centrifugal constant $D^\ast$ as a function of $B^\ast$  in comparison with the experimental data listed in Ref. \cite{Callegari_2000, GrebenevOCS, Nauta_2001, Nauta2001, Lindsay_2005, Choi2006, Raston_2011, Raston_2012, Morrison_2013, Raston_2014, Faulkner_2018, chatterley_rotational_2020}.  In agreement with the established experimental and theoretical result, $D^\ast$ measured in helium droplets is found to be $10^2-10^4$ times larger than the corresponding gas-phase value. The light rotors with $B \gtrsim \SI{3}{cm^{-1}}$, whose  $B/B^*$ ratio is barely distinguishable from 1, show large $D^\ast$ of the order of $\SI{0.01}{cm^{-1}}$  only if the interaction  parameter, $u$, is large (orange and blue lines). Otherwise, $D^\ast$ does not scale with $B^\ast$ (green line) and might be comparable to the gas-phase centrifugal constant for some of the molecules. In the case of heavy rotors this tendency is not apparent, $D^*$ shows persistent dependence on $B^*$ in a wide range of $u$. In particular, one can see that, similarly to \autoref{eq:strong_constants}, the scaling of $D^\ast$ closely resembles the already mentioned empirical formula, $D^* = 0.031 \times B^{* 1.818}$, found in Ref. \cite{Choi2006}.

\section{Conclusions}
\label{sec:conclusions}

Thus we presented a simple quantum mechanical model describing the rotational level structure of molecules in superfluid helium nanodroplets and, in particular, capturing highly excited states (recent work\cite{CherepanovPRA21} compared the model calculations with experiment up to $J \gtrsim 15$ for I$_2$ and CS$_2$ molecules).  Here we provided details on the theoretical machinery of the model, benchmarked its results against the data on the effective spectroscopic constants $B^*$ and $D^*$ for a broad range of molecules. Although the model is already based on a simplified version of the previously reported angulon Hamiltonian~\cite{SchmidtLem15}, we have substantially simplified it further and have shown that several   properties of molecules in superfluids can be understood by analyzing a simple $2 \times 2$ matrix, \autoref{eq:tls}.

Among other results, we gathered the following insights:

(i) A linear molecule in superfluid He can be described as an effective symmetric top, with an additional quantum number describing the projection of superfluid angular momentum on the molecular $z$-axis. Coupling between the superfluid and molecular rotational angular momenta is reminiscent of that between the electronic and rotational angular momenta in the gas-phase radicals, such as OH or NO. Analyzing different possible angular momentum coupling schemes in terms of ``many-body Hund's cases'' would be very interesting to do in the future.  

(ii) Analyzing the structure of such a symmetric top, whose states can be mixed by molecule--helium interactions, furnishes a few qualitative insights. For example, the crossover between the rotational behavior of light and heavy molecules in a superfluid (approximately at $B\sim2-3$~cm$^{-1}$) can be explained in terms of the many-particle wavefunction structure shown in \autoref{fig:renorm}(c), which, in turn, follows from the $L$-dependent energy gap $\Delta_L$ shown in \autoref{fig:band}(a).

The results presented here and in Ref.~\cite{CherepanovPRA21} reveal that the structure of the highly excited rotational states can substantially deviate from the gas-phase-like \autoref{eq:B*D*}, in particular for heavier molecules, such as I$_2$ and CS$_2$. This deviation needs to be taken into account while creating molecular superrotors using the optical centrifuge technique~\cite{Centrifuge99, KorobenkoPRL14}. In particular, one might need to redefine the adiabaticity criterion of molecule-laser interactions and to use non-linear ramp pulses in order to account for the threshold of the states as shown in \autoref{fig:band}.

\begin{acknowledgments}
I.C.~acknowledges the support by the European Union's Horizon 2020 research and innovation programme under the Marie Sk\l{}odowska-Curie Grant Agreement No.~665385. G.B.~acknowledges support from the Austrian Science Fund (FWF), under project No.~M2461-N27 and from the Deutsche Forschungsgemeinschaft (DFG, German Research Foundation) under Germany's Excellence Strategy EXC2181/1-390900948 (the Heidelberg STRUCTURES Excellence Cluster). M.L.~acknowledges support by the Austrian Science Fund (FWF), under project No.~P29902-N27, and by the European Research Council (ERC) Starting Grant No.~801770 (ANGULON).  H.S. acknowledges support from the Independent Research Fund Denmark (Project No. 8021-00232B) and from the Villum Fonden through a Villum Investigator Grant No. 25886.
\end{acknowledgments}


\bibliographystyle{apsrev4-1}
\bibliography{References}

\end{document}